\documentclass{evn2024}
\usepackage{aas_macros}
\usepackage{natbib}

\usepackage{graphics,graphicx}
\usepackage{caption}
\usepackage{subcaption}
\usepackage{paralist}
\usepackage{multicol}

\usepackage{amsmath,amssymb,amsfonts,latexsym,cancel}
\usepackage{sidecap}
\usepackage{mdwlist}
\usepackage{bibentry}
\usepackage{threeparttable}
\usepackage{booktabs}
\usepackage{epstopdf}
\usepackage[percent]{overpic}
\usepackage[utf8]{inputenc} 
\usepackage{orcidlink}

\usepackage{hyperref}
\hypersetup{
  colorlinks   = true, 
  urlcolor     = blue,
  linkcolor    = blue, 
  citecolor    = blue, 
  filecolor    = black 
}

\providecommand{\BIBand}{\&}

\usepackage{paralist}
\let\olditem\item
\renewenvironment{thebibliography}[1]{
  \noindent
  \let\par\relax
  \renewcommand{\item}[1][]{\olditem}
  \inparaenum[{[}1{]}]}{\endinparaenum}
  
\def\rg{{R$_{\rm g}$}}

\def\microarc{$\mu$arcsec}

\usepackage{txfonts}

\newcommand{\phrvd}{Physical Review D}
\newcommand{\apaa}{Acquisition, Processing and Archiving of Astronomical Images}

\newcommand{\natas}{Nature Astronomy}

\newcommand{\msun}{\ensuremath{M_{\scriptscriptstyle \odot}}}

\begin{document}

\title{Demographics of black holes at $<$100 R$_{\rm g}$ scales: accretion flows, jets, and shadows}

        \author
        {
        \normalsize{
        Dhanya G. Nair~\orcidlink{0000-0001-5357-7805}\inst{1,2}\thanks{Corresponding author \email{dhanyagnair01@gmail.com}}
        \and
        Neil M. Nagar~\orcidlink{0000-0001-6920-662X}\inst{1}
        \and
        Venkatessh Ramakrishnan~\orcidlink{0000-0002-9248-086X}\inst{3,40,1}
        \and
        Maciek Wielgus~\orcidlink{0000-0002-8635-4242}\inst{4}
        \and
        Vicente Arratia~\orcidlink{0000-0003-4785-2297}\inst{1}
        \and
        Thomas P. Krichbaum~\orcidlink{0000-0002-4892-9586}\inst{2}
        \and
        Xinyue A. Zhang~\orcidlink{0009-0007-5412-1894}\inst{5,8}
        \and
        Angelo Ricarte~\orcidlink{0000-0001-5287-0452}\inst{5,8}
        \and
        Silpa S.~\orcidlink{0000-0003-0667-7074}\inst{1}
        \and
        Joaqu\'in Hern\'andez-Y\'evenes~\orcidlink{0000-0001-5845-7538}\inst{1}
        \and
        Nicole M. Ford~\orcidlink{0000-0001-8921-3624}\inst{6} 
        \and
        Bidisha Bandyopadhyay~\orcidlink{0000-0002-2138-8564}\inst{1}
        \and  
        Mark Gurwell~\orcidlink{0000-0003-0685-3621}\inst{5}
         \and
        Roman Burridge~\orcidlink{0009-0000-3265-7726}\inst{12} 
        \and
        Dominic W. Pesce~\orcidlink{0000-0002-5278-9221}\inst{5,8}
        \and
        Sheperd S. Doeleman~\orcidlink{0000-0002-9031-0904}\inst{5,8}
        \and
        Jae-Young Kim~\orcidlink{0000-0001-8229-7183}\inst{63,2}
        \and
        Daewon Kim~\orcidlink{0000-0003-4997-2153}\inst{2}
        \and
        Michael Janssen~\orcidlink{0000-0001-8685-6544}\inst{19,2}
        \and
        Sebastiano D. von Fellenberg~\orcidlink{0000-0002-9156-2249}\inst{2}
        \and
        Christian M. Fromm~\orcidlink{0000-0002-1827-1656}\inst{21,20,2} 
        \and
        Deokhyeong Lee~\orcidlink{0009-0003-2122-9437}\inst{27} 
        \and
        Heino Falcke~\orcidlink{0000-0002-2526-6724}\inst{19}
        \and
        Jan Wagner~\orcidlink{0000-0003-1105-6109}\inst{2}
        \and 
        Geoffrey C. Bower~\orcidlink{0000-0003-4056-9982}\inst{11,12}
        \and
        Anne-Kathrin Baczko~\orcidlink{0000-0003-3090-3975}\inst{10,2}
        \and
        Dong-Jin Kim~\orcidlink{0000-0002-7038-2118}\inst{61}
        \and
        Kazunori Akiyama~\orcidlink{0000-0002-9475-4254}\inst{7,8,9}
        \and
        Keiichi Asada~\orcidlink{0000-0001-6988-8763}\inst{28}
        \and
        Patricia Arevalo~\orcidlink{0000-0001-8503-9809}\inst{43,44}
        \and
        Hayley Bignall~\orcidlink{0000-0001-6247-3071}\inst{62,61} 
        \and
        Lindy Blackburn~\orcidlink{0000-0002-9030-642X}\inst{5,8}
        \and
        Avery E. Broderick~\orcidlink{0000-0002-3351-760X}\inst{13,14,15}
        \and
        Andreas Brunthaler~\orcidlink{0000-0003-4468-761X}\inst{2}
        \and
        Chi-kwan Chan~\orcidlink{0000-0001-6337-6126}\inst{16,17,18}
        \and
        Akihiro Doi\inst{60}
        \and
        Vincent L. Fish~\orcidlink{0000-0002-7128-9345}\inst{7}
        \and
        Edward Fomalont~\orcidlink{0000-0002-9036-2747}\inst{42}
        \and
        Jos\'e L. G\'omez~\orcidlink{0000-0003-4190-7613}\inst{4}
        \and
        Daryl Haggard~\orcidlink{0000-0001-6803-2138}\inst{6,22}
        \and
        Kazuhiro Hada~\orcidlink{0000-0001-6906-772X}\inst{23,24}
        \and
        Rodrigo Herrera-Camus~\orcidlink{0000-0002-2775-0595}\inst{1} 
        \and
        Daniel Hoak~\orcidlink{0000-0001-5126-8048}\inst{7}
        \and
        David Hughes\inst{57}
        \and
        Julie Hlavacek-Larrondo~\orcidlink{0000-0001-7271-7340}\inst{45}
        \and
        Svetlana Jorstad~\orcidlink{0000-0001-6158-1708}\inst{25}
        \and
        Michael D. Johnson~\orcidlink{0000-0002-4120-3029}\inst{5,8}
        \and
        Tomohisa Kawashima~\orcidlink{0000-0001-8527-0496}\inst{52}
        \and
        Garrett K. Keating~\orcidlink{0000-0002-3490-146X}\inst{5}
        \and
        Preeti Kharb~\orcidlink{0000-0003-3203-1613}\inst{26}
        \and
        Jun Yi Koay~\orcidlink{0000-0002-7029-6658}\inst{29}
        \and
        Shoko Koyama~\orcidlink{0000-0002-3723-3372}\inst{28,29}
        \and
        Cheng-Yu Kuo~\orcidlink{0000-0001-6211-5581}\inst{53,28}
        \and
        Nathan W. C. Leigh~\orcidlink{0000-0003-0347-276X}\inst{1}
        \and
        Paulina Lira~\orcidlink{0000-0003-1523-9164}\inst{1,43}
        \and
        Michael Lindqvist~\orcidlink{0000-0002-3669-0715}\inst{10}
        \and
        Andrei P. Lobanov~\orcidlink{0000-0003-1622-1484}\inst{2}
        \and
        Wen-Ping Lo~\orcidlink{0000-0003-1869-2503}\inst{28,56}
        \and
        Ru-Sen Lu ~\orcidlink{0000-0002-7692-7967}\inst{31,32,2}
        \and
        Sera Markoff~\orcidlink{0000-0001-9564-0876}\inst{33,34}
        \and
        Nicholas R. MacDonald~\orcidlink{0000-0002-6684-8691}\inst{64,2}       
        \and
        Mary Loli Mart\'{i}nez-Aldama~\orcidlink{0000-0002-7843-7689}\inst{1}
        \and
        Lynn D. Matthews~\orcidlink{0000-0002-3728-8082}\inst{7}
        \and
        Satoki Matsushita~\orcidlink{0000-0002-2127-7880}\inst{28}
        \and
        Mar Mezcua~\orcidlink{0000-0003-4440-259X}\inst{46,47}
        \and
        Monika Moscibrodzka~\orcidlink{0000-0002-4661-6332}\inst{19}
        \and
        Hendrik M\"{u}ller~\orcidlink{0000-0002-9250-0197}\inst{42,2}
        \and
        Hiroshi Nagai~\orcidlink{0000-0003-0292-3645}\inst{9,24}
        \and
        Masanori Nakamura~\orcidlink{0000-0001-6081-2420}\inst{51,28}
        \and
        Priyamvada Natarajan~\orcidlink{0000-0002-5554-8896}\inst{48,49,8}
        \and
        Gopal Narayanan~\orcidlink{0000-0002-4723-6569}\inst{35}
        \and
        Michael A. Nowak~\orcidlink{0000-0001-6923-1315}\inst{36}
        \and
        H\'ector Ra\'ul Olivares S\'anchez~\orcidlink{0000-0001-6833-7580}\inst{55}
        \and
        Jongho Park~\orcidlink{0000-0001-6558-9053}\inst{37}
        \and
        Dimitrios Psaltis~\orcidlink{0000-0003-1035-3240}\inst{38}
        \and
        Hung-Yi Pu~\orcidlink{0000-0001-9270-8812}\inst{58,59,28}
        \and
        Oliver Porth~\orcidlink{0000-0002-4584-2557}\inst{33,20}
        \and
        Ramprasad Rao~\orcidlink{0000-0002-1407-7944}\inst{5}
        \and
        Cormac Reynolds~\orcidlink{0000-0002-8978-0626}\inst{61}
        \and
        Rodrigo Reeves~\orcidlink{0000-0001-5704-271X}\inst{1} 
        \and
        Cristina Romero-Ca\~nizales~\orcidlink{0000-0001-6301-9073}\inst{28}
        \and
        Eduardo Ros~\orcidlink{0000-0001-9503-4892}\inst{2}
        \and
        Helge Rottmann~\orcidlink{0000-0003-1799-8228}\inst{2}
        \and
        Alan L. Roy~\orcidlink{0000-0002-1931-0135}\inst{2}
        \and
        Dominik Schleicher~\orcidlink{0000-0002-9642-120X}\inst{1}
        \and
        Tuomas Savolainen~\orcidlink{0000-0001-6214-1085}\inst{39,40,2}
        \and
        C. M. Violette Impellizzeri~\orcidlink{0000-0002-3443-2472}\inst{41,42}
        \and
        Ezequiel Treister~\orcidlink{0000-0001-7568-6412}\inst{50}
        \and
        Kaj Wiik~\orcidlink{0000-0002-0862-3398}\inst{54,3,40}
        \and
        J. Anton Zensus~\orcidlink{0000-0001-7470-3321}\inst{2}  
        }
        }

\authorrunning{D.~G.~Nair et al.}
\titlerunning{Demographics of black holes at $<$100 R$_{\rm g}$ scales: accretion flows, jets, and shadows}

\abstract{
\normalsize
Using the Event Horizon Telescope (EHT), the gravitationally lensed rings around the supermassive black holes (SMBHs) in Messier 87 (M87), which has billions of solar masses and a strong jet, and Sagittarius A* (Sgr A*), which has millions of solar masses and a weak or no jet, have now been successfully imaged at a resolution under 10 gravitational radii (\( R_{\rm g} = GM/c^2 \), where \(M\) is the SMBH mass). To expand studies beyond M87 and Sgr A*, we have constructed the 
`Event Horizon and Environs (ETHER)' sample, a comprehensive database currently 
encompassing approximately 3.15 million SMBH mass estimates, $\sim$ 20,000 Very-Long Baseline Interferometry (VLBI) radio flux densities, and $\sim$ 36,000 hard X-ray flux densities.  
This database is designed to identify and optimize target selection for the 
EHT and its upgrades on the ground and in space. Within the ETHER database, we have identified a `Gold Sample' (GS) of nearby low-luminosity Active Galactic Nuclei (AGNs) that are optimal for studying jet bases, and potentially imaging black hole shadows, with the EHT and its upgrades. 

We observed 27 of these AGNs using the EHT from 2022 to 2024, providing a unique opportunity to resolve and image  accretion flows and jets at resolutions of $\leq$ 100 \rg. Only a few SMBHs have sufficiently high enough flux density to be imaged at scales of $\leq$ 50 \rg with the present EHT. 
Among these are M87, Sgr A*, NGC\,4594 (Sombrero/M104), NGC\,4261, and NGC\,4374 (Messier 84/M84). Of these, NGC\,4261, Sombrero, and M84 have been observed and/or are scheduled for deep imaging with EHT+ALMA from 2023 to 2025. Sombrero, NGC\,4261, M84, NGC\,4278, and NGC\,5232 are clearly detected in our EHT+ALMA observations in 2022, indicating that the 230 GHz flux density from the accretion flows is significantly high. Ongoing imaging of the ETHER GS will enable measurements of black hole mass and spin, help constrain General Relativity (GR), and enrich our understanding of jet launching and accretion inflows across a broad multi-parameter space, including black hole mass, spin, accretion rate, and orientation.    
}

\maketitle
\section{Introduction}
\vspace{-0.2em} 
Supermassive black holes (SMBHs, $\sim10^5-10^{10}$ \msun) are likely present in most galaxies, with masses $\sim$0.5\% of the host galaxy bulge \citep[e.g.,][]{Saglia2016}.
SMBHs and their host galaxies evolve together, with Active Galactic Nuclei (AGNs) accretion and feedback 
impacting the host galaxy. The Event Horizon Telescope (EHT) now provides direct evidence for the existence of SMBHs and the associated accretion flows \citep{EHT1,EHT1Sgr}.

A SMBH is characterized by three independent external observables: mass, spin, and charge. 
SMBHs at high Eddington accretion rates typically host an optically-thick, geometrically-thin accretion disk extending from 10s to 1000s of \rg. At low Eddington accretion rates (characteristic of low-luminosity AGNs, or LLAGNs), they exhibit a quasi-spherical optically-thin accretion inflow extending down to the innermost tens of \rg~\citep[e.g.,][]{Narayan1996}. Bipolar relativistic plasma jets launched from within 100 \rg~of the SMBH are more common in the latter. 
 Several jet launching mechanisms have been proposed \citep[e.g.,][]{Blandford1977, Blandford1982}, with the rotational energy extraction either from the accretion flow or the SMBH spin. The physics of jet launching and collimation remains an active area of research \citep[e.g.,][]{Broderick2015b, Tchekhovskoy2015,EHT1} that is primarily limited by the paucity of observational data on scales less than hundreds of \rg.

 When an SMBH is enveloped in an optically-thin accretion inflow and/or has an emitting jet base, we expect to see: 
(1) the accretion inflow and jet \citep[e.g.,][]{Johnson2023,Janssen2021} down to 10s of \rg;
(2) a bright ring produced by gravitational lensing around the SMBH (diameter 
$\sim$10.4--10.8 \rg, almost independent of black hole spin and orientation \citealt{Falcke2000,Johannsen2012,Gralla2019});
(3) a deficit of emission from inside the ring, or the 'shadow', due to photons lost to the event horizon or deflected away from the line of sight \citep{Falcke2000,Johnson2023}. 
The shadow and ring shapes tightly constrain the SMBH mass, spin, space-time metric, and `no hair' theorem \citep[e.g.,][]{Broderick2014, Vigeland2011}. 
The accretion inflow's morphology and polarization can constrain accretion disk physics and help discriminate between disk states \citep[e.g., magnetization and temperature;][]{EHT5,EHT2024}.
Meanwhile, the jet's morphology, distance from the SMBH, magnetic field strength and orientation \citep[from polarization measurements, e.g.,][]{Fish2013,Issaoun2022,Paraschos2024} and lateral profile \citep[e.g.,][]{Asada2012,Hada2013b,Walker2018} provide critical constraints on current jet theory and numerical simulations \citep{Savolainen2008, OSullivan2009,Nakamura2018}.

\begin{figure}[t]
\centerline{\includegraphics[width=0.5\textwidth]{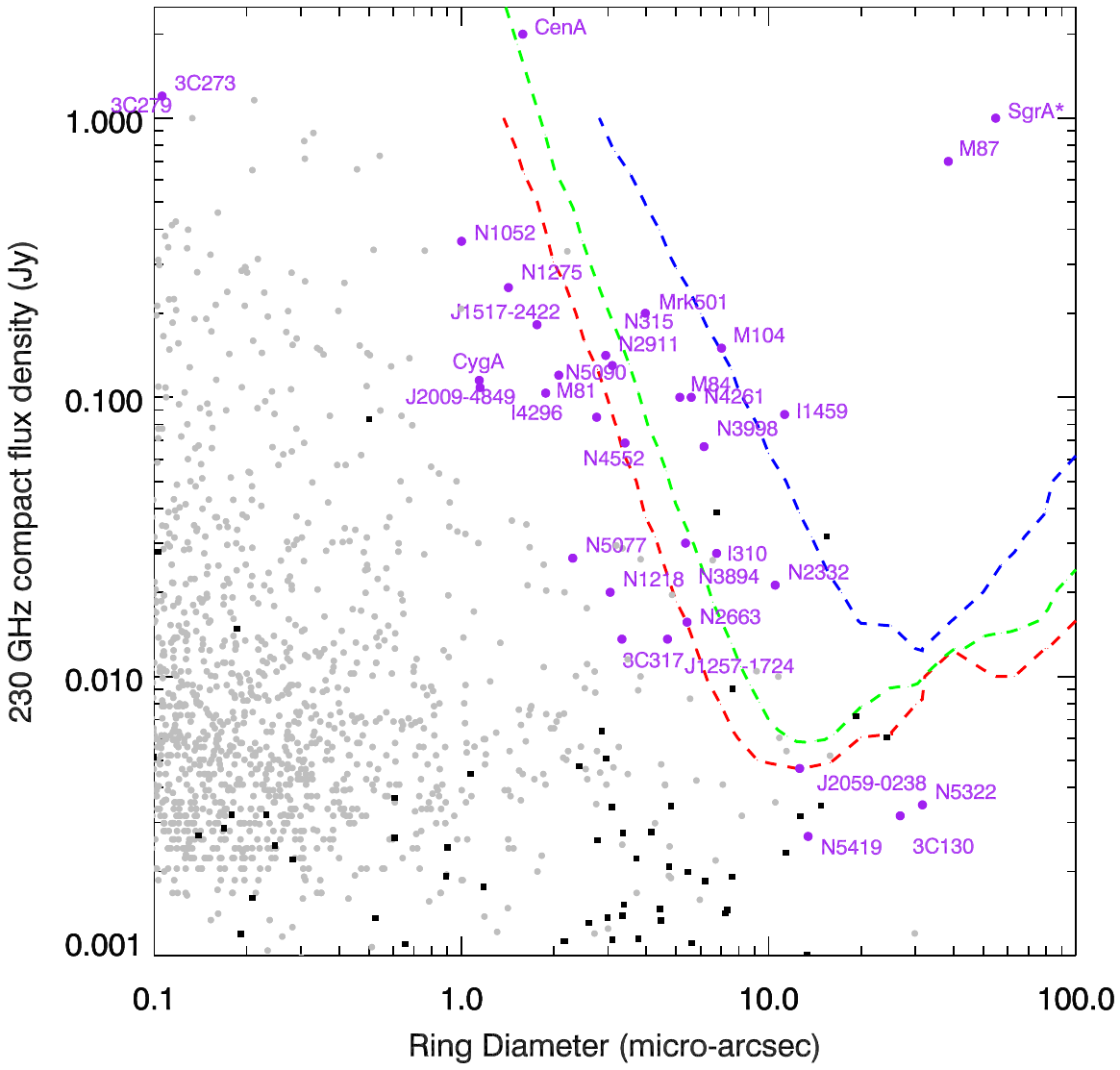}}
\caption{
The expected 230 GHz flux density in the EHT field of view
as a function of the diameter of the gravitationally lensed ring around the SMBH 
(here 10.4 \rg) in the ETHER sample. Circles denote VLBI-detected SMBHs, with `GS' sources highlighed in purple. Black squares denote other SMBHs detected in our SMA and ACA 230 GHz observation runs
(see Sect.~\ref{sec:230 GHz survey with ALMA Compact Array Submillimeter Array}, Nair et al., in prep.).
The ngEHT is expected to measure the SMBH masses, spins and shadows above
the red, green, and blue lines, respectively \citep{Pesce2022, Johnson2023}. 
}  
\label{figring}
\end{figure}

\begin{figure*}[ht!]
\centerline{
    \begin{subfigure}{0.28\textwidth}
        \centering
            \caption{}
            \includegraphics[width=\textwidth]
            {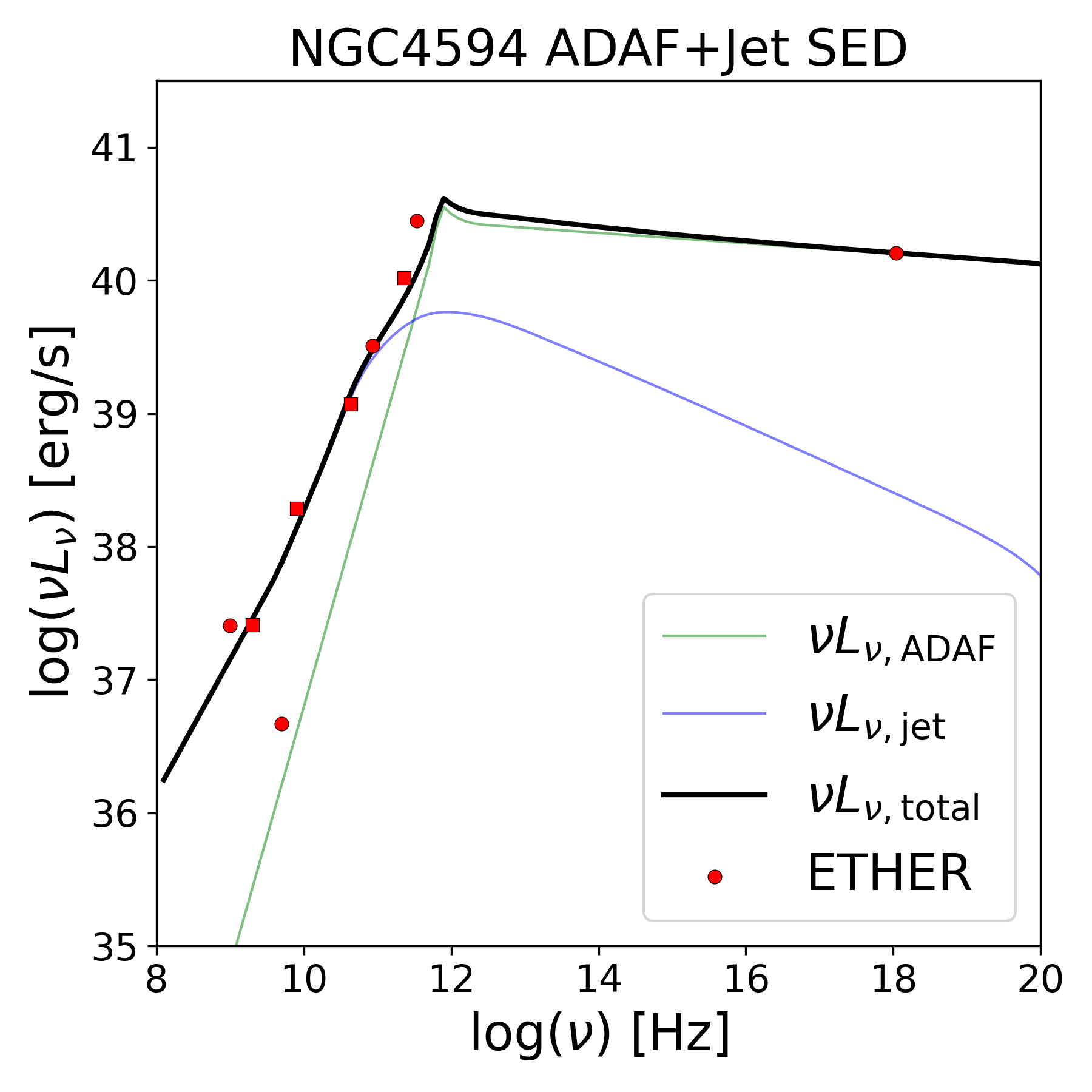}
    \end{subfigure}
    \hfill
    \begin{subfigure}{0.28\textwidth}
        \centering
            \caption{}
            \includegraphics[width=\textwidth]
            {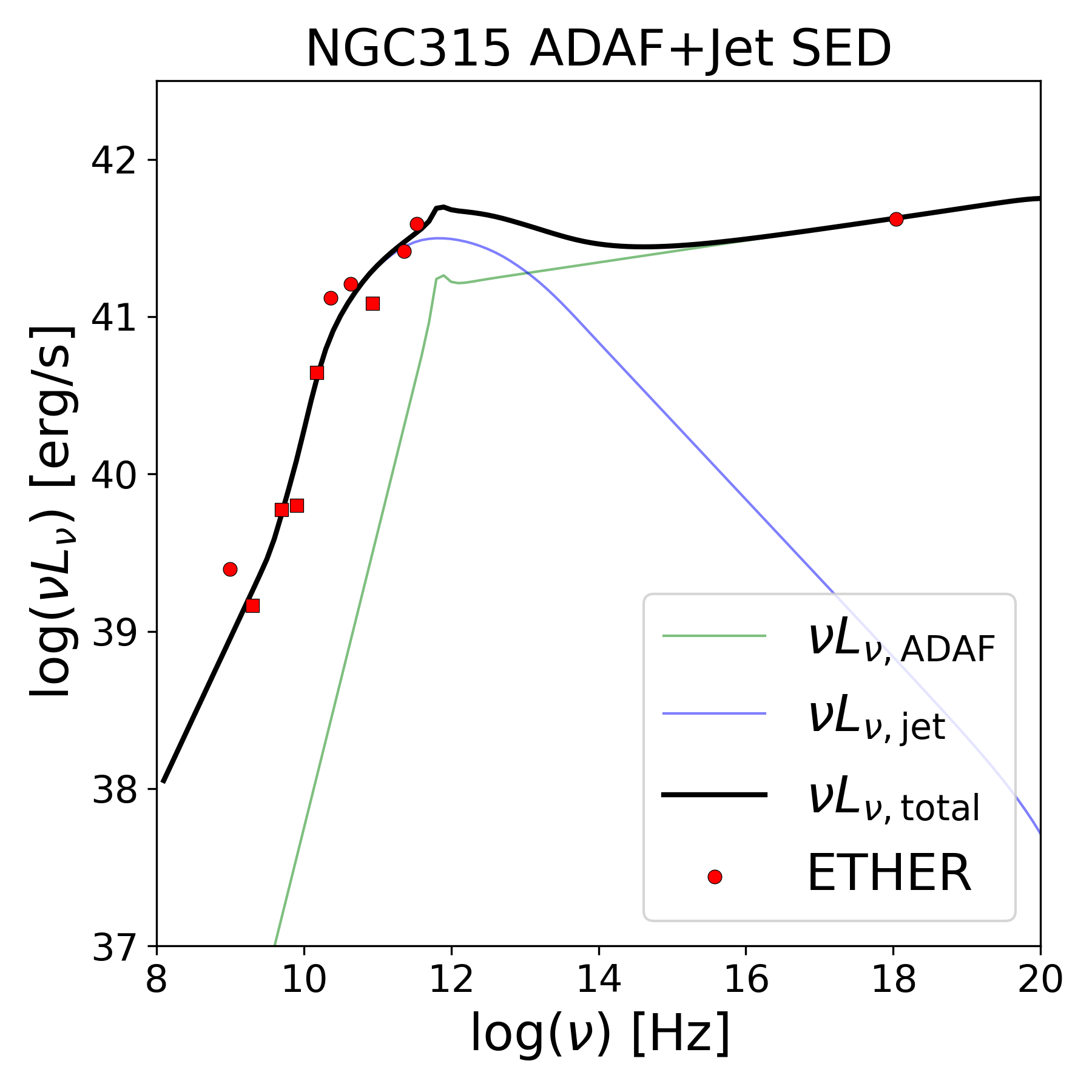}
    \end{subfigure}
    \hfill
    \begin{subfigure}{0.44\textwidth}
        \centering
            \caption{}
           \begin{tabular}{|c|c|c|c|c|}
            \hline
             Name    &  $D$    &  $\log$       & Ring        & VLBI:            \\
                     &  [Mpc]  &  $M_{BH}$     & size        & [Peak-Frequency   \\
                     &         &  [$\msun$]    & [$\mu as$]  & -Resolution]      \\  \hline
            IC1459   & 25.90  &  9.45$^a$   & 11.27   &  56$^g$, 43, 0.5    \\ \hline  
            NGC4594  &  9.87  &  8.82$^b$       & 7.03   & 91$^h$, 43, 0.5   \\ 
                     &        &  9.00$^c$       & 10.41  &                   \\ \hline 
            M84      & 18.50  &  8.96$^a$      &  5.14  & 86$^i$, 43, 0.5    \\ 
                     &        &  9.18$^d$      &  8.35  &                   \\ \hline
            NGC4261  & 31.10  &  9.22$^e$   & 5.60  & 59$^j$, 86, 0.4   \\ \hline
            NGC1218  & 113.79  & 9.52$^f$  & 3.05   & 113$^g$, 43, 0.5     \\ \hline 
        \end{tabular}
    \small\em{a - \cite{Ruffa2024}; 
    b - \cite{Gultekin2019};
    c - \cite{Kormendy1996};
    d - \cite{Bower1998};
    e - \cite{Boizelle2021};
    f - \cite{Bosch2015};
    g - Nair et al., in prep.;
    h - \cite{Hada2013a};
    i - \cite{Wang2022};
    j -  \cite{Middelberg2005}.  
    }
        \label{tab:table1}
    \end{subfigure}
           }    
    \caption{
    \normalsize{    
    Left (a \& b): Illustrative examples of simple analytic ADAF+jet fits (black curve) to the nuclear, highest resolution available (red points, or red squares if VLBI) radio and hard X-ray data
    for two ETHER galaxies: one ADAF-dominated (NGC\,4594 (a)) and the other jet-dominated (NGC\,315 (b)), as discussed in Sect~\ref{sec:The sub-mm `bump' and ADAF+Jet models to predict EHT fluxes}.   
    The 43 GHz VLBI flux densities and morphologies presented in Sect.~\ref{sec:ETHER: 43 GHz VLBA Imaging} (see e.g., in Fig.\ref{LLAGN-NGC1218}) are crucial for constraining the model jet parameters and enabling more detailed modeling.    \\
    Right (c): A table showing the properties of a few selected large-ring SMBHs in the `ETHER GS', as described in Sect.~\ref{The Event Horizon and Environs (ETHER) Sample}, which we observed with EHT+ALMA, GMVA+ALMA, and the 43 GHz VLBA. Columns: 1~--~Name; 2~--~Angular distance in Megaparsec (Mpc); 3~--~Logarithm (base 10) of black hole mass in solar masses (\msun); 4~--~Diameter of the gravitationally lensed ring around SMBH in \microarc; 5~--~The peak flux density (mJy), observation frequency (GHz), and resolution (milliarcseconds) from previous VLBI imaging in the literature or from our 43 GHz VLBA imaging (PID: BN065B, BN065C).
    }}
    \label{figadaf}
\end{figure*}

The Event Horizon Telescope (EHT) array is a global VLBI network of radio telescopes operating at 230 GHz \citep{EHT2},
which has imaged SMBH shadows in M87 \citep{EHT1} and Sgr A* \citep{EHT1Sgr}, thus testing GR and constraining black hole mass \citep{EHT6,EHTSgr6}. It has also imaged the inner jets of several other SMBHs, such as 3C 279 \citep{Kim2020}, Centaurus A \citep{Janssen2021}, J1924-2914 \citep{Issaoun2022}, NRAO 530 \citep{Jorstad2023}, and 3C84 \citep{Paraschos2024}, 
but all at resolutions coarser than 100 \rg.
The Global Millimeter VLBI Array (GMVA) is a global VLBI network of radio telescopes operating at 86 GHz frequency \citep{Krichbaum2006, Boccardi2017}.
The GMVA+ALMA and EHT+ALMA achieve native resolutions (FWHM) of $\sim$45 \microarc\ at 86~GHz \citep{Boccardi2017,Nair2019} and $\sim$20 \microarc\ at 230~GHz \citep{EHT1}, respectively. 
Using `super-resolution' imaging techniques \citep{Chael2016,Akiyama2017}, the resolutions can reach as low as $\sim$ 15 \microarc\ with the GMVA and $\sim$ 8\,\microarc\ with the EHT, if the SNR and (u,v) coverage are sufficient.
In the high-SNR regime, Bayesian imaging and modeling techniques \citep{Broderick2020a,Broderick2020b} enable precise estimations of model parameters, such as ring diameters for horizon-scale sources and distances between moving features \citep[e.g.,][]{Kim2020}, enhancing intrinsic resolutions by factors of approximately 3 to 5. Additionally, recent advancements in imaging techniques \citep[e.g.,][]{Muller2022,Kim2024} continue to improve these capabilities.
The EHT can now detect sources as faint as 30--50 mJy \citep{EHT3}, with next-generation upgrades to the EHT (ngEHT) expected to lower this limit to a few to 10 mJy \citep{Doeleman2023}. 
This opens the possibility to apply the transformational results in M87 \citep{EHT1} and Sgr~A$^*$ \citep{EHT1Sgr} to a sample of SMBHs with a broader range of masses, spins, accretion rates, and orientations by imaging the inner hundreds of \rg.

\vspace{-1.0em}
\section{Results \& discussion}
\vspace{-0.1em}
\subsection{\textbf{Event Horizon and Environs (ETHER) sample:}}
\label{The Event Horizon and Environs (ETHER) Sample}
Simulations suggest that the EHT with ngEHT expansion, utilizing super-resolution techniques, is expected to measure $\sim$ 50 SMBH masses, $\sim$ 30 SMBH spins, and $\sim$ 7 SMBH shadows \citep{Pesce2022, Johnson2023}.
As noted by \cite{Johnson2023}, these mass measurements will provide the most precise SMBH masses for a sample to date. Spin measurements (particularly for low-Eddington luminosity SMBHs) may help trace SMBH merger and accretion history \citep[e.g.,][]{Ricarte2023, Ricarte2024}, meanwhile, shadow measurements offer strong tests of gravity.
To identify this elusive `Gold Sample' (GS) and fully leverage the capabilities of the EHT and ngEHT, we have compiled the Event Horizon and Environs (ETHER) sample. This extensive dataset includes black hole mass measurements and estimates, radio to hard X-ray flux densities, and spectral energy distribution (SED) information from catalogs, literature, and our own observations.

Currently, the ETHER database consists of $\sim$ 3.15 million SMBH mass estimates,
$\sim$ 20,000 VLBI radio flux densities, and $\sim$ 36,000 hard X-ray flux densities. 
An additional $\sim$ 210,000 SMBHs have some radio flux density measurement at frequencies between 1.4 and 345 GHz.
This database and its algorithms, along with data source references, were first presented in \citet{Ramakrishnan2023} and updated in \citet{Hernandez2024}.
ETHER is continuously evolving, with the incorporation of a wider literature and new observational programs (Silpa S. et al., in prep.).

Currently, there are only $\sim$ 244 black hole mass measurements from methods such as stellar dynamics \citep[e.g.,][]{Thater2019,Liepold2020},
gas kinematics \citep[e.g.,][]{Onishi2017,Kuo2011,Nguyen2021,Boizelle2019,Boizelle2021}, 
reverberation mapping \citep[RM;][]{Bentz2015},
ETHER thus primarily depends on black hole mass estimations, with $\sim$ 240,000 via the M-sigma relation \citep{Saglia2016}, 
$\sim$ 2.7 million via the M-$L_\mathrm{bulge}$ relation primarily from \cite{Hernandez2024},
524,000 from single-epoch spectroscopy of broad line AGNs using RM-based relationships
\citep[e.g.,][]{Dalla2020,Rakshit2021},
and several thousand from other fundamental plane type relationships.

ETHER contains predictions for the 230 GHz and 345 GHz flux densities in the EHT field of view for $\sim$ 40,000 SMBHs. 
These are obtained via: 
(a) extrapolations of observed VLBI (mas-scale) 5 to 86 GHz flux densities for $\sim$ 20,000 sources \citep[e.g., from][the Radio Fundamental Catalog and our new Very-Long Baseline Array (VLBA) observations from Nair et al., in prep.; note that we extrapolated VLBI measurements from the highest observed frequency available to 230 GHz, applying a typical spectral index of -0.4 and incorporating a median flux attenuation factor to adjust for higher resolution at increased frequencies (varying from 0.4--0.8 for 5--86 GHz)]{Nair2019, Cheng2020, Lister2018, Doi2016, Helmboldt2007};
(b) extrapolations of arcsec-scale 230 GHz flux densities
from our ALMA Compact Array (ACA) \& Submillimeter Array (SMA) surveys (Nair et al., in prep.), the ALMA calibrator database, and literature;
and (c) predictions from Advection Dominated Accretion Flow (ADAF) plus jet model fitting to the radio and hard X-ray flux densities for a given SMBH mass measurement or estimate (Arratia et al., in prep.). 

Combining all of this information, we have now identified an initial `GS' from the ETHER dataset, comprising the largest SMBH rings with the brightest estimated EHT flux densities from their jets and accretion flows. 
The most recent data for the VLBI-detected sources in the ETHER
sample are displayed in Fig.~\ref{figring}, featuring several GS targets.
The GS spans a wide range of SMBH masses 
($\log M_{\rm{BH}} = 7.74-9.72 \, M_{\odot}$), 
Eddington rates ($\log L_{\rm{Edd}} = -6.59 $  to $-1.92$), jet powers, and morphologies, with details of a few selected top targets given in panel (c) of Fig.~\ref{figadaf}.
We are observing these targets with ALMA, SMA, VLBA, GMVA, and EHT to leverage their combined capabilities at high angular resolution for further study, with details and results discussed in the following subsections: Sect.~\ref{sec:230 GHz survey with ALMA Compact Array Submillimeter Array},~\ref{sec:ETHER: 43 GHz VLBA Imaging}, ~\ref{ETHER: GMVA observations overview}, and \ref{ETHER: EHT observations overview}.

\begin{figure*}[ht!]
\centerline{
            \includegraphics[width=1.0\textwidth]{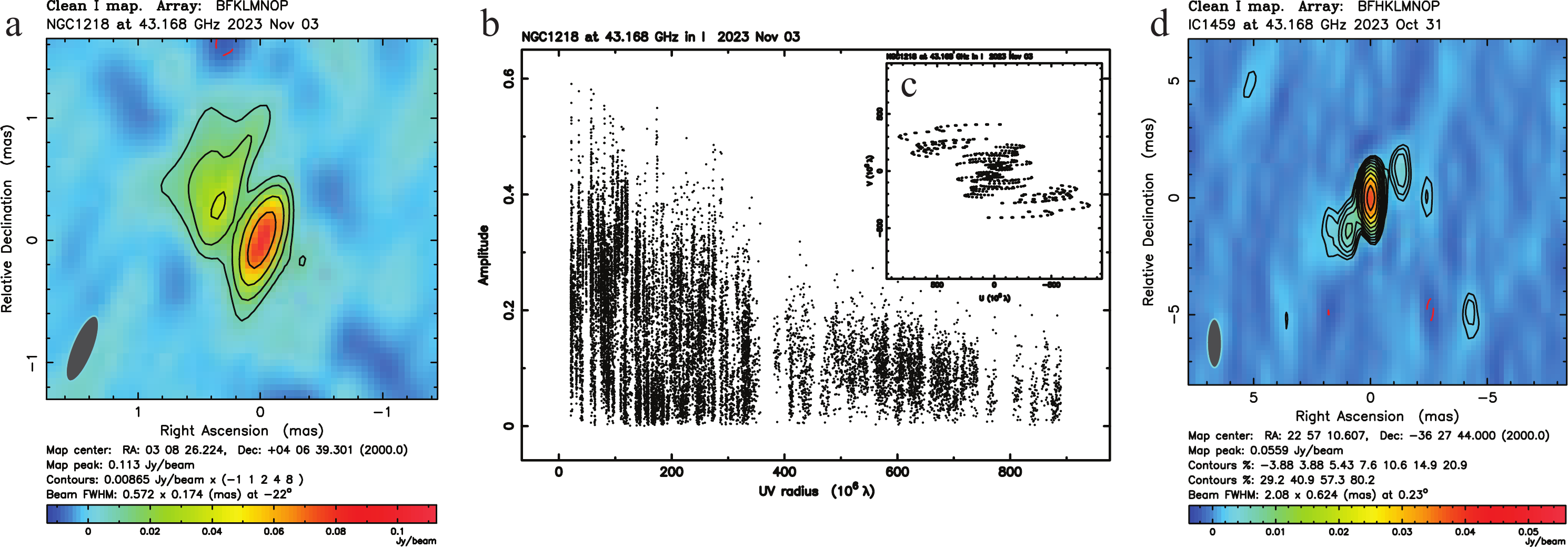}
           }         
\caption{Example results from our VLBA 43 GHz imaging of ETHER targets (Sect.~\ref{sec:ETHER: 43 GHz VLBA Imaging}) include: (a) the NGC~1218 image (total clean flux: $257 \pm 3 $ mJy, peak flux: 113 mJy), (b) NGC~1218's $uv$ visibility amplitude vs. $uv$ radius, (c) NGC~1218's $uv$ coverage, and (d) the IC~1459 image (total clean flux: $65.2 \pm 1.2$ mJy, peak flux: 55.9 mJy) (from Nair et al., in prep.).
}           
\label{LLAGN-NGC1218}
\end{figure*}

\vspace{-1.0em}
\subsection{\textbf{SED modeling to predict EHT flux densities}}
\label{sec:The sub-mm `bump' and ADAF+Jet models to predict EHT fluxes}
\vspace{-0.3em}
Nuclear radio emission in LLAGNs originates primarily from two sources: 
the outflowing jet,
which typically dominates at cm wavelengths, and the accretion inflow, which is typically stronger at mm and sub-mm wavelengths. This leads to the ``sub-mm bump," an increase in sub-mm emission beyond what cm-wave extrapolation predicts \citep[e.g.,][]{Doi2005,Bandyopadhyay2019}. 
To model this phenomenon, we utilize detailed \citep[e.g.,][]{Bandyopadhyay2019} or analytic (\citealt{Lucchini2022,Spada2001}) plus ADAF models \citep{Pesce2021}.
Our broadband SED models, first introduced in \cite{Ramakrishnan2023} and further developed by Arratia et al. (in prep.), can predict the expected EHT flux density from both the jet and ADAF inflow.
Examples of the ADAF+jet models are shown in Fig.~\ref{figadaf}.
The modeling and fitting processes require key inputs including distance, redshift, estimated SMBH mass, hard X-ray flux densities, and cm/mm-wave radio flux densities ideally from VLBI observations.
Approximately 36,000 ETHER sources are obtained from the \textit{Chandra} Source Catalog, NED, and/or the \textit{eROSITA} eRASS1 release \citep{Merloni2024}.
Integrating these with radio-flux densities from our VLBA observations (Sect. \ref{sec:ETHER: 43 GHz VLBA Imaging}), supplemented by our 5~arcsec 230 GHz flux densities from ALMA-ACA and SMA campaigns (Sect.~\ref{sec:230 GHz survey with ALMA Compact Array Submillimeter Array}), is vital for ADAF$+$jet SED fitting and selecting the GS.

Accurate SED modeling for ETHER targets depends on having high-resolution and simultaneous radio and X-ray data. 
Our recent and upcoming high-resolution \textit{NuSTAR} and high-cadence \textit{NICER} 
X-ray programs (PIDs: \href{https://heasarc.gsfc.nasa.gov/docs/nustar/ao8/c8_acceptarg.html}{8229} and \href{https://heasarc.gsfc.nasa.gov/docs/nustar/ao10/c10_accepted_program_abstracts.html}{10244})
targeting multiple GS galaxies (e.g., Sombrero and NGC\,4261), will enable us to build broad, time-resolved SEDs, fit our accretion and jet models, and predict EHT band flux densities.


\begin{figure*}[ht!]
\centerline{
    \begin{subfigure}{0.25\textwidth}
        \centering
            \caption{}
            \includegraphics[width=\textwidth]{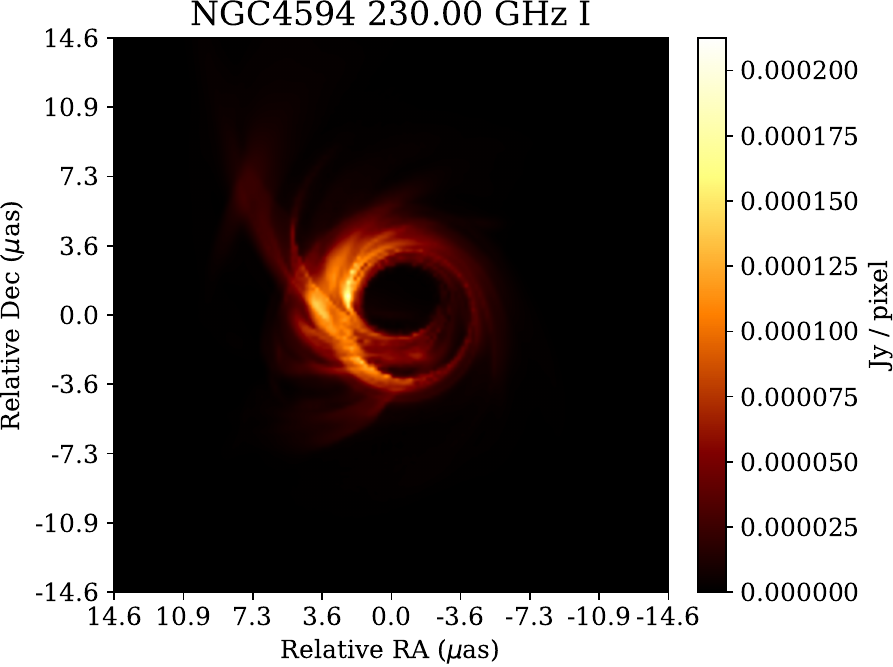}
    \end{subfigure}
    \hfill
    \begin{subfigure}{0.25\textwidth}
        \centering
            \caption{}
            \includegraphics[width=\textwidth]{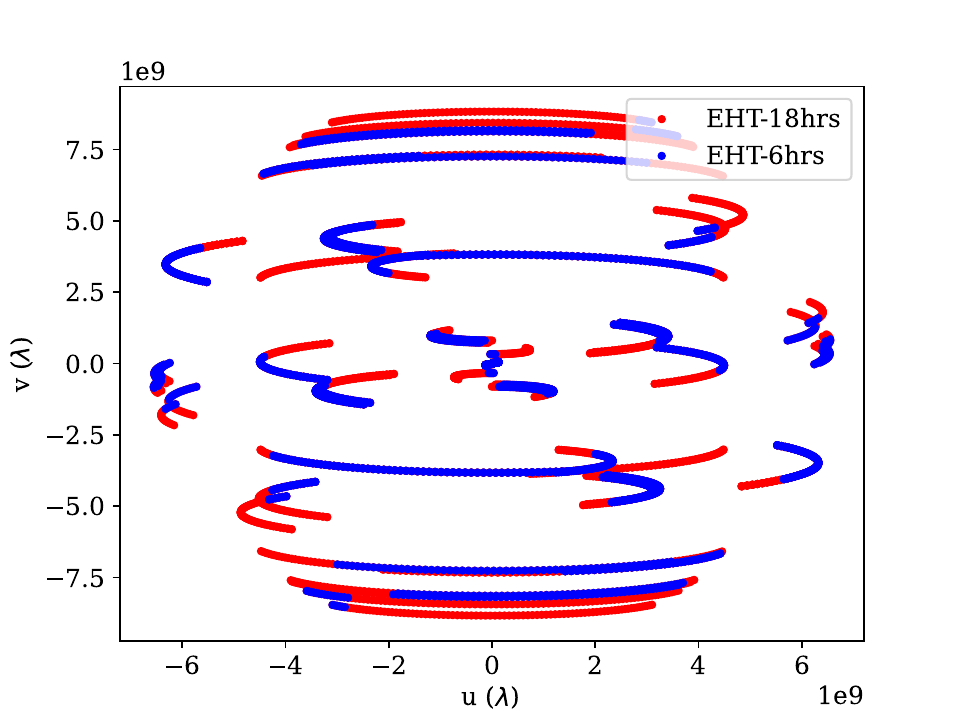}
    \end{subfigure}
    \hfill
    \begin{subfigure}{0.25\textwidth}
        \centering
            \caption{}
            \includegraphics[width=\textwidth]{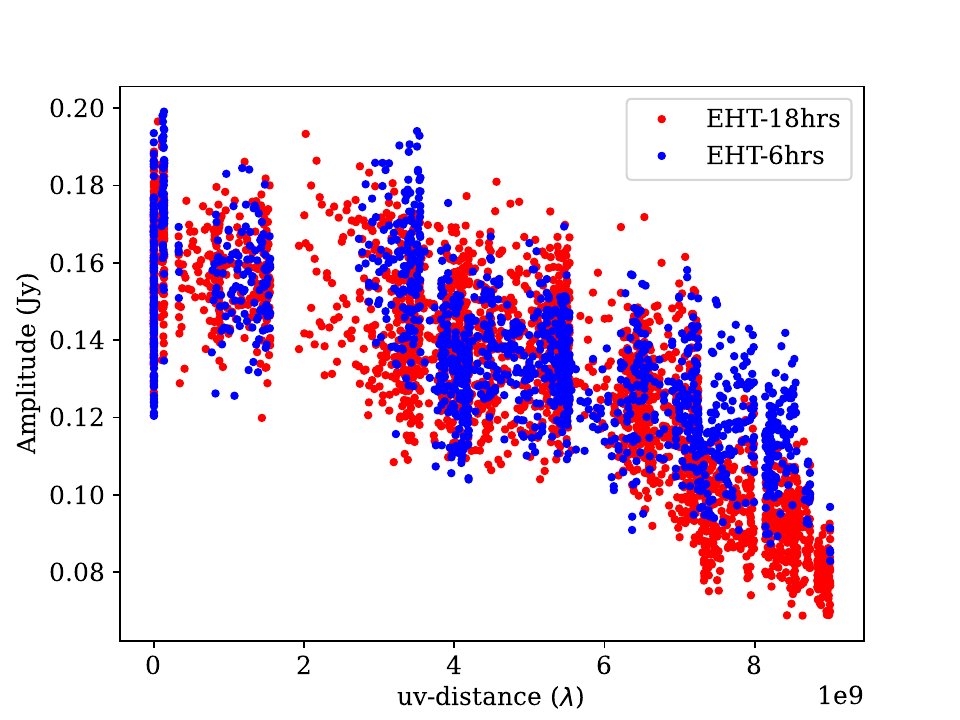}
    \end{subfigure}
    \hfill
    \begin{subfigure}{0.21\textwidth}
        \centering
            \caption{}
            \includegraphics[width=\textwidth]{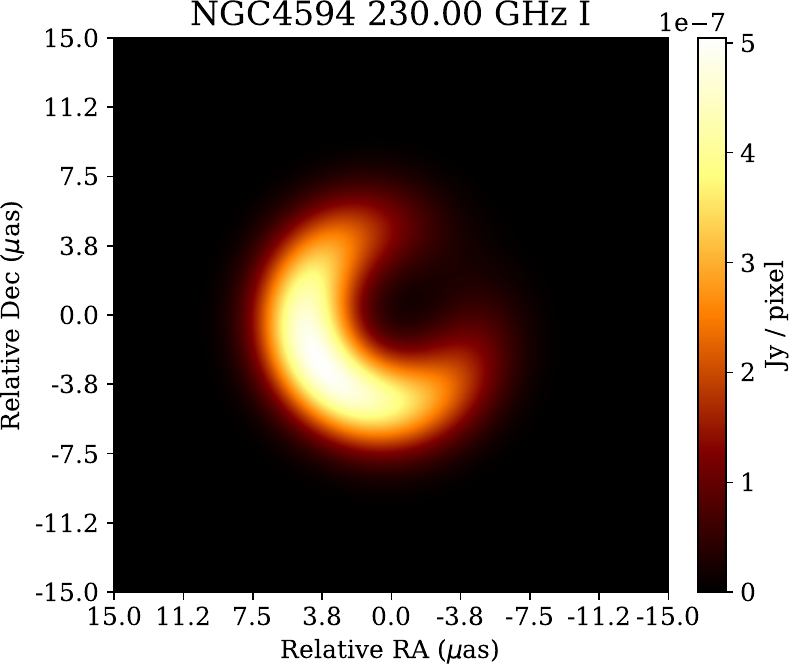}
    \end{subfigure}
           }
\caption{
\normalsize{
(a): Ray-traced General Relativistic Magneto-hydrodynamic (GRMHD) simulations representing the center of the Sombrero galaxy, scaled to the mean total 230 GHz flux density of Sombrero (198 mJy), its mass \((M = 10^{8.82} \, M_\odot)\), distance \((D = 9.87 \, \text{Mpc})\), expected inclination \((50^\circ)\) and with the spin/jet position angle being vertical (from \citealt{Zhang2024}). 
(b): The simulated $uv$ coverage expected from our EHT observations of Sombrero in April 2024 (6 hours) is shown in blue. For reference, the maximal $uv$ coverage obtainable by the present EHT for Sombrero is that shown in blue and red together. 
(c): Visibility amplitudes obtained for the synthetic observation of Sombrero as a function of $uv$ distance, with a random gain amplitude corruption of 10\%. 
Data from the 6-hour track is shown in blue, while the maximal obtainable dataset is shown in blue and red. 
(d): Best-fit model to the maximal synthetic data (blue plus red in panels (b) and (c)), obtained by fitting the GRMHD model image with a thick ring of total intensity = $160$ mJy
(ring diameter = $9.94 \pm 0.04$ \microarc, 
ring thickness (FWHM of Gaussian convolution) = $4.22 \pm 0.08$ \microarc,
azimuthal brightness variations in the ring determined by a Fourier mode expansion with 
complex Fourier coefficients (beta1\_{abs} = $0.48 \pm 0.003$, 
beta1\_{arg} =   $-124.02^\circ \pm 0.17^\circ $).
The $\chi^2$ values for closure phases and closure amplitudes are 2.82 and 1.56, respectively.
The jet emission is very weak, so the source appears as a very compact thick ring with a diameter of $\sim 10$ \microarc\, (from Nair et al., in prep.).
}
}  
\label{Sombrero-grmhd}
\end{figure*}


\begin{figure*}[ht!]
\centerline{
    \begin{subfigure}{0.23\textwidth}
        \centering
            \caption{}
            \includegraphics[width=\textwidth]{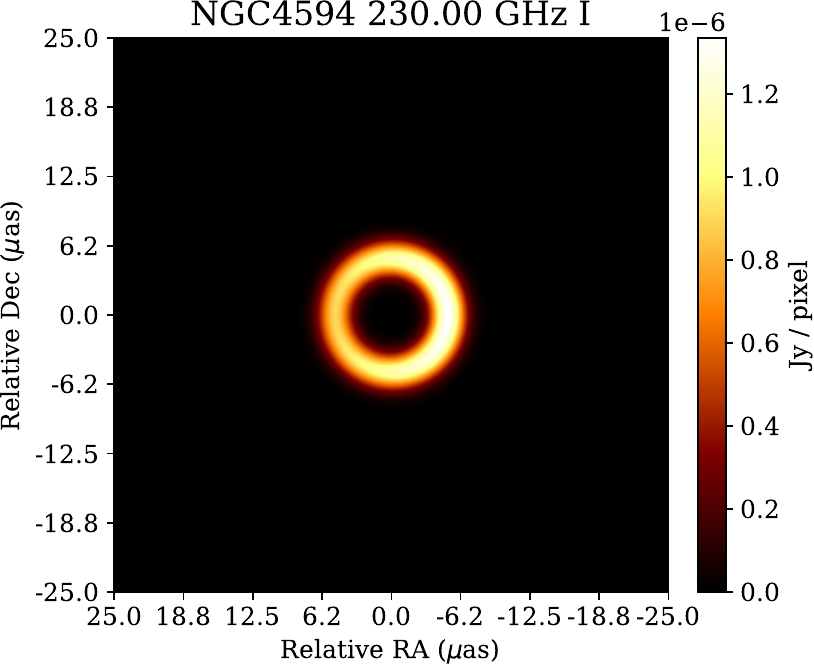}
    \end{subfigure}
    \hfill
    \begin{subfigure}{0.25\textwidth}
        \centering
            \caption{}
            \includegraphics[width=\textwidth]{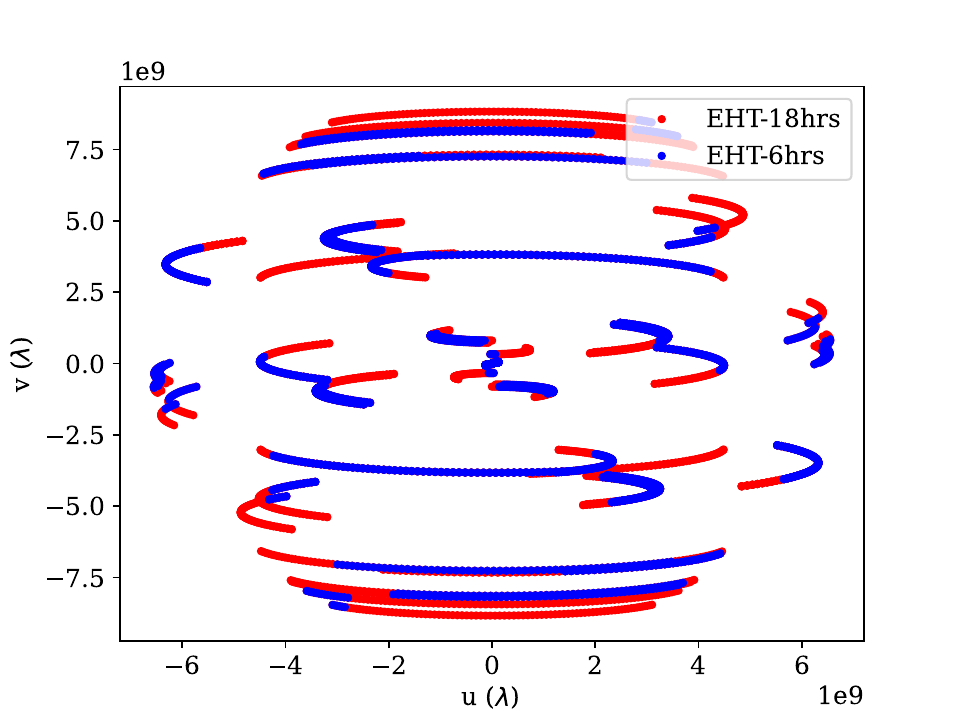}
    \end{subfigure}
    \hfill
    \begin{subfigure}{0.25\textwidth}
        \centering
            \caption{}
            \includegraphics[width=\textwidth]{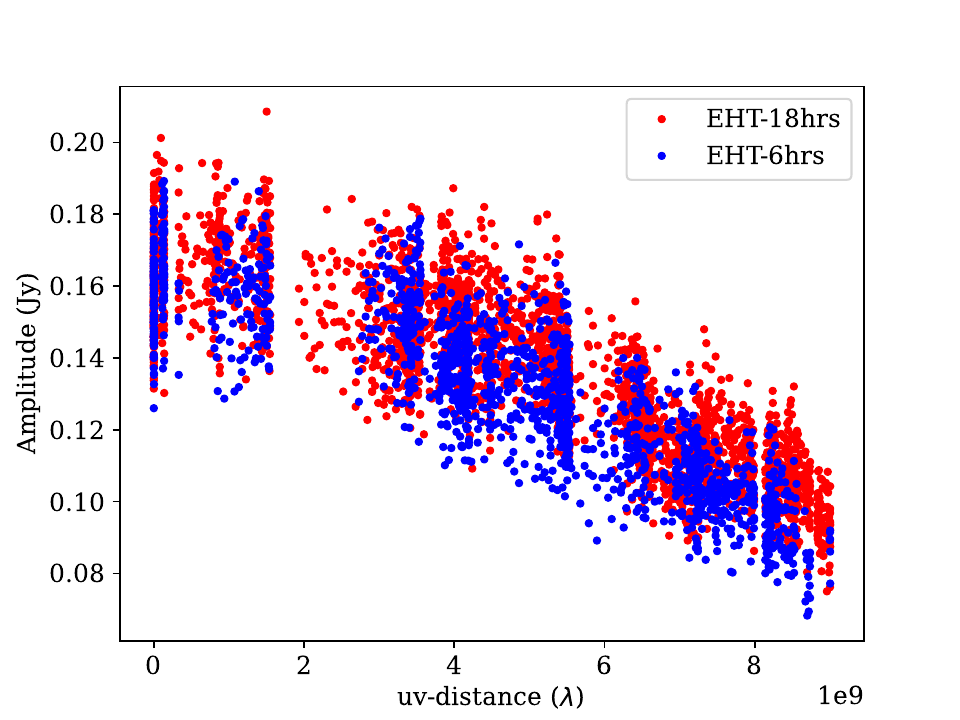}
    \end{subfigure}
    \hfill
    \begin{subfigure}{0.23\textwidth}
        \centering
            \caption{}
            \includegraphics[width=\textwidth]{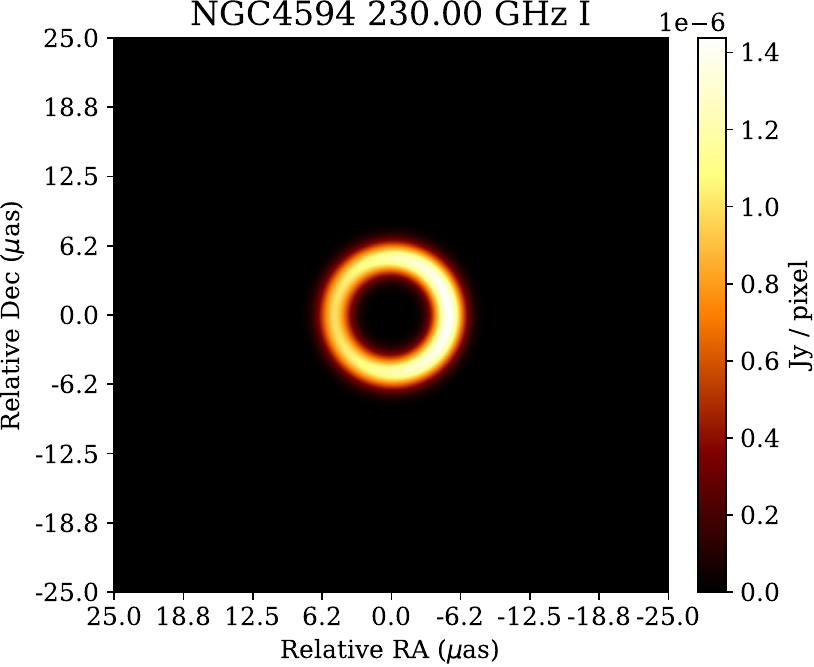}
    \end{subfigure}
           }
\caption{
\normalsize{ 
(a): A geometric model representing the center of the Sombrero galaxy as a thick ring
(total intensity = 160 mJy, ring diameter = 10.41 \microarc, 
ring thickness = 2.60 \microarc, azimuthal brightness variations in the ring determined by a Fourier mode expansion with complex Fourier coefficients [beta1\_{abs} = $0.10 $, beta1\_{arg} =  $90^\circ $]).
(b): The simulated $uv$ coverage expected from our EHT observations of Sombrero in April 2024 (6 hours) is shown in blue. For reference, the maximal $uv$ coverage obtainable by the present EHT for Sombrero is that shown in blue and red together.   
(c): Visibility amplitudes obtained for the synthetic observation of the geometric model of Sombrero as a function of $uv$ distance, with a random gain amplitude corruption of 10\%. Data from the 6-hour track is shown in blue, while the maximal obtainable dataset is shown in blue and red. 
(d): Best-fit solution obtained by fitting a geometric model of the source with a thick ring of total intensity = $160$ mJy (ring diameter = $10.49 \pm 0.09$ \microarc, ring thickness = $2.25 \pm 0.38$ \microarc, azimuthal brightness variations in the ring determined by a Fourier mode expansion with complex Fourier coefficients [beta1\_{abs} = $0.093 \pm 0.006 $, beta1\_{arg} =  $88.04^\circ \pm 2.05^\circ $], 
to the maximal synthetic data (blue plus red in panels (b) and (c)).
The $\chi^2$ values for closure phases and closure amplitudes are 0.98 and 0.99, respectively (from Nair et al., in prep.).}
}
\label{Sombrero-synthetic}
\end{figure*}

\begin{figure*}[ht!]
\centerline{
    \begin{subfigure}{0.35\textwidth}
        \centering
            \caption{}
            \includegraphics[width=\textwidth]{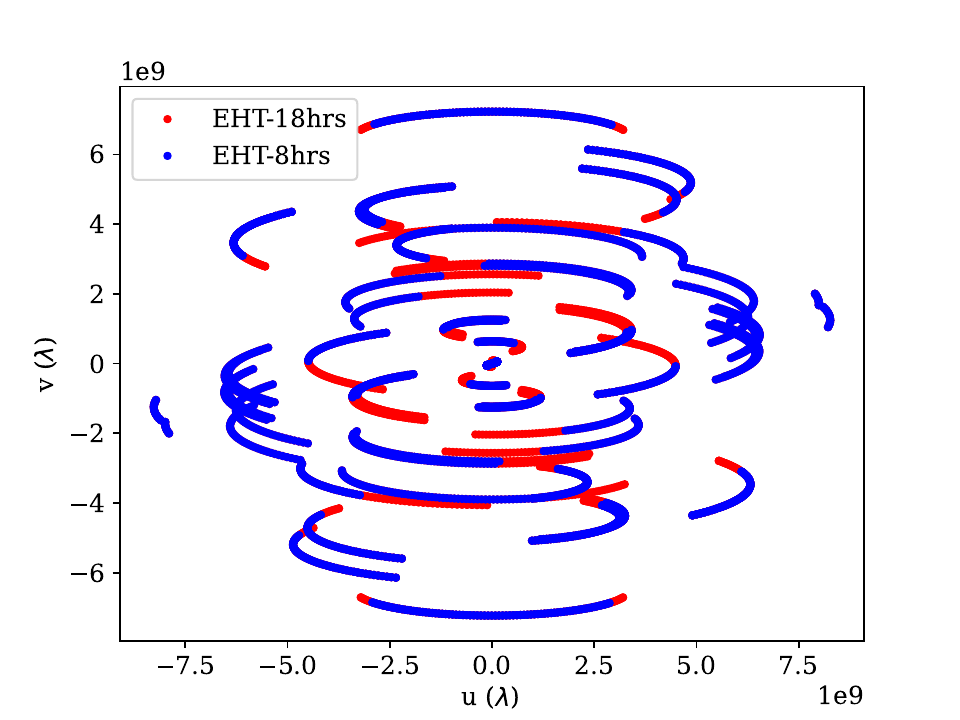}
    \end{subfigure}
    \hfill
    \begin{subfigure}{0.35\textwidth}
        \centering
            \caption{}
            \includegraphics[width=\textwidth]{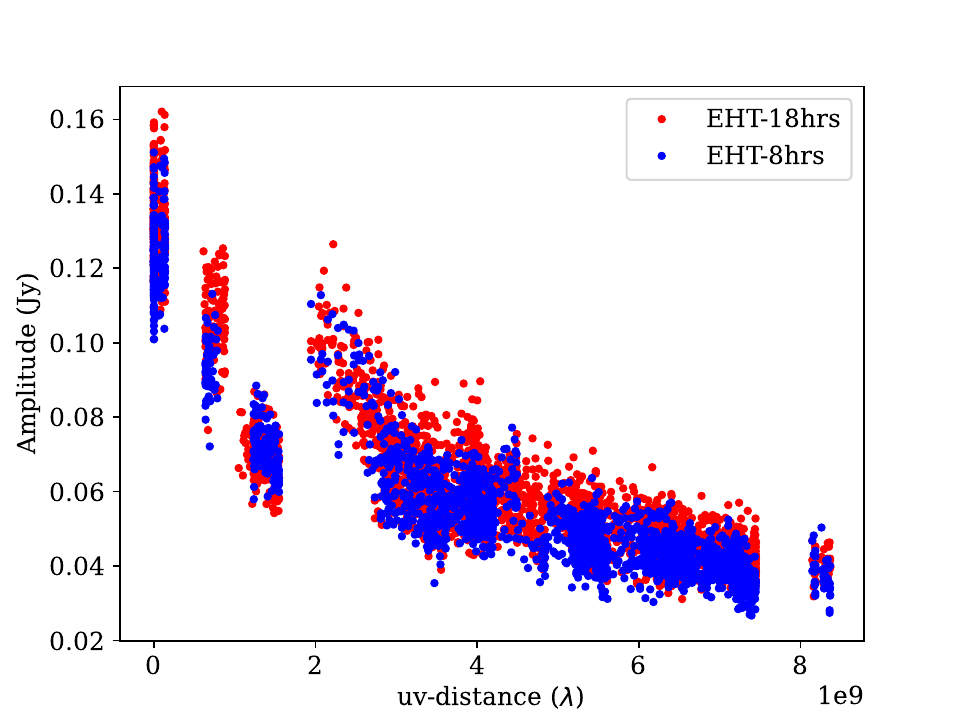}
    \end{subfigure}
    \hfill
    \begin{subfigure}{0.35\textwidth}
        \centering
            \caption{}
            \includegraphics[width=\textwidth]{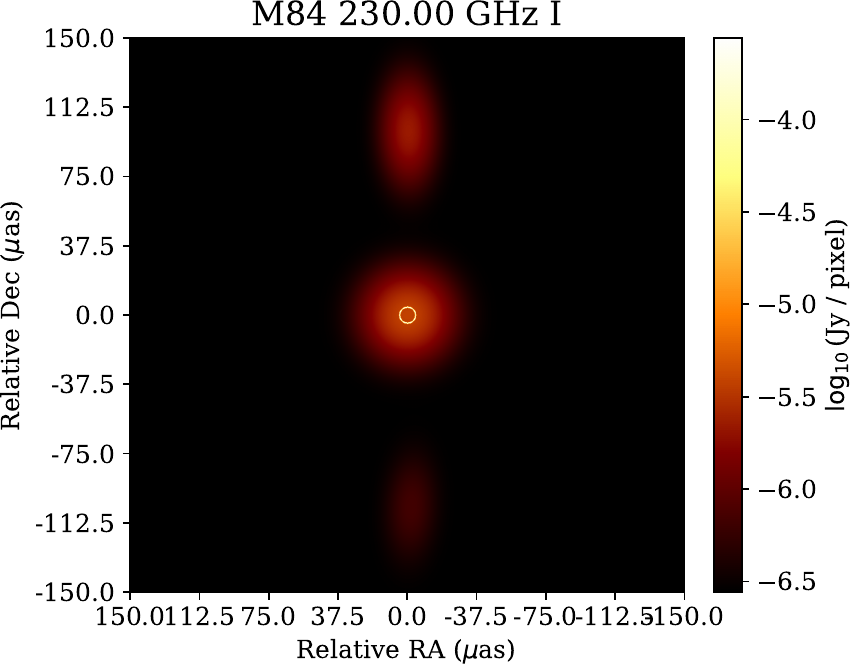}
    \end{subfigure}
           }
    \caption{
\normalsize{
(a): The simulated $uv$ coverage of the EHT expected for our approved 8-hour track to observe M84 in April 2025 is shown in blue. For reference, the maximal $uv$ coverage obtainable by the present EHT for M84 is that shown in blue and red together.  
(b): Visibility amplitudes as a function of $uv$ distance obtained for the synthetic observation of M84, with a random gain amplitude corruption of 10\%. 
Data from the 8-hour track is shown in blue, while the maximal obtainable dataset is shown in blue and red. 
(c): Best-fit model for the maximal synthetic data (blue plus red in panels (a) and (b)) of M84,
obtained by fitting a geometric model with an 8.5\,\microarc\, ring of 50\,mJy; 
a circular Gaussian representing accretion inflow (49.2 mJy, FWHM = 32.4\,\microarc); 
an elliptical Gaussian representing the jet base (20.6 mJy, FWHM\(_{\text{major}}\) = 39.7\,\microarc, FWHM\(_{\text{minor}}\) = 19.9\,\microarc, PA = 0.7\(^\circ\)
x0 = -0.51\,\microarc, y0 = 99.4\,\microarc); 
and an elliptical Gaussian representing the counter-jet base (4.2\,mJy, FWHM\(_{\text{major}}\) = 40.4\,\microarc, FWHM\(_{\text{minor}}\) = 19.3\,\microarc, PA = 177.0\(^\circ\), x0 = -1.87\,\microarc, y0 = -103.5\,\microarc). 
The $\chi^2$ values for closure phases and closure amplitudes are 1.12 and 0.98,
respectively. The morphology of the assumed synthetic model is fully and accurately recovered (from Nair et al., in prep.).
}   
}
\label{M84-synthetic}
\end{figure*}


\vspace{-1.0em}
\subsection{\textbf{ETHER: 230 GHz survey with ALMA Compact Array \& Submillimeter Array}}
\label{sec:230 GHz survey with ALMA Compact Array Submillimeter Array}

We conducted an extensive 230 GHz survey of a complete sample of SMBHs with estimated ring sizes $>$ 3.0 \microarc\ to constrain their morphologies and measure peak 230 GHz flux densities. Together with the ADAF and ADAF$+$jet modeling presented in Sec.~\ref{sec:The sub-mm `bump' and ADAF+Jet models to predict EHT fluxes}, this will lead to better predictions of the EHT flux.
This survey was carried out using the Atacama Compact Array (ACA), where 
929 sources in southern declinations $<$ $20^\circ$ were observed for a total of 86 hours during ALMA Cycles 8 and 9 (\href{https://almascience.eso.org/observing/highest-priority-projects}{PIDs: 2021.1.00966.S, 2022.1.00049.S}),
and 10 (\href{https://almascience.eso.org/observing/highest-priority-projects}{PID: 2023.1.01099.S}) from 2021 to 2024, and using the Submillimeter Array (SMA), where 96 sources in northern declinations $>$ $20^\circ$ were observed over 24 hours (PID: 2022B-S022) from 2022 to 2023. This effort is crucial for identifying `ADAF-only' sources (i.e., without cm-wave jets) for inclusion in the GS.

We achieved a 10\% detection rate with the ACA and a 15\% detection rate with the SMA as demonstrated in Fig.~\ref{figring}, identifying potential sources for the EHT or ngEHT (Nair et al., in prep.). For example, NGC 5872 and PGC 1086306 were detected with peak flux densities of $12.7 \pm 3.3$ mJy and $11.7 \pm 1.6$ mJy,
respectively, using ACA, while NGC 5353 and NGC 2146 were detected with peak flux densities of $20.8 \pm 2.7$ mJy and $15.6 \pm 3.1$ mJy, 
respectively, using SMA. The 5$\sigma$ upper limits were around 8--10 mJy at ALMA and SMA, indicating that even among current non-detections, there could be interesting science that could be done with the ngEHT.
  
The observed and EHT-predicted 230 GHz flux densities constitute an important driver of the technical requirements of the ngEHT. Our initial results suggest that SMBH demographic studies with the ngEHT would highly benefit from the capability of imaging $\sim$ 10 mJy SMBHs and detecting $\sim$ 1 mJy SMBHs in $\sim$ 1 hour of integration.

\vspace{-1.0em}
\subsection{\textbf{ETHER: 43 GHz VLBA Imaging}}
\label{sec:ETHER: 43 GHz VLBA Imaging} 
\vspace{-0.3em}
The VLBA at 43 GHz is the highest frequency sensitive phase-referenced VLBI network,
and coincidentally, this frequency is near the transition between jet dominance and (potential) ADAF dominance in radio emission, and thus VLBA 43 GHz imaging can provide valuable constraints for our SED modeling discussed in Sect.~\ref{sec:The sub-mm `bump' and ADAF+Jet models to predict EHT fluxes}. We thus observed 32 of our top `GS' ETHER targets, including 23 galaxies with estimated ring sizes $\geq$ 3 \microarc\ and nine with estimated ring sizes $\geq$ 1 \microarc that lacked VLBI observations at $\nu \geq 43$ GHz but were detected in previous VLBI experiments at lower frequency (typically 8.4 GHz).

Through several VLBA programs (PIDs: BR222, BN054, BN065B, BN065C) between 2018-2023, 
we have observed and detected all ETHER targets with the 43 GHz VLBA so far (Nair et al., in prep.). The data are calibrated in AIPS \citep{Greisen1990} 
or using the CASA rPICARD pipeline \citep{Janssen2019,EHT3} for various epochs and then imaged using the Caltech DIFMAP software package \citep{Shepherd1994}.
The measured peak flux densities of these sources range from 27 to 155 mJy, and the targets show a variety of 43 GHz nuclear morphologies, including core-twin-jet, core-jet, and compact core structures. Two examples, NGC\,1218 and IC\,1459, are shown in Fig.~\ref{LLAGN-NGC1218}, along with typical $uv$ coverage, and a typical $uv$ visibility amplitude plot. We found several flat spectrum nuclei 
underlining the need for 43 GHz VLBA imaging of all large (even if fainter) rings previously detected by lower frequency VLBI. The success of our 43 GHz VLBA detections of these large ring SMBHs directly motivated EHT+ALMA snapshot/imaging observations of these sources in April 2022, 2023, and 2024 (see Sect.~\ref{ETHER: EHT observations overview} for details). 

\vspace{-1.0em}
\subsection{\textbf{ETHER: 86 GHz GMVA+ALMA observations}}
\label{ETHER: GMVA observations overview}
\vspace{-0.3em}
The largest SMBH rings in the `ETHER GS', with expected sizes ranging from 0.6 to 8.9 \microarc, and 43/15 GHz VLBI core detections of flux densities ranging from 40 to 416 mJy (as found in Sect.~\ref{sec:ETHER: 43 GHz VLBA Imaging} and the literature), were observed at 86 GHz with GMVA+ALMA in May 2023 and October 2024.
The 2023 program included Sombrero, M84, IC\,1459, and NGC\,3998 (GMVA-only) for deep-imaging, along with fringe tests of NGC\,4261, NGC\,4278, and NGC\,5077 (flux densities: 91 to 390 mJy at 43/15 GHz; PID: MN004 \& 2022.1.00366.V, 19 hrs). 
The 2024 GMVA-only program, including the APEX telescope, observed NGC\,1218, 3C\,218, NGC\,3894, UGC\,2783, NGC\,6061, NGC\,6251, ICRF\,J152122.5$+$0420, NGC\,5232, and NGC\,4552 (flux densities: 40 to 416 mJy at 43/15 GHz; PID: MN005, 6.75 hrs), thus completing the characterization of all galaxies currently imageable at $\leq$ 100 \rg\ with the current EHT. The forthcoming data will determine the 86 GHz nuclear morphologies and flux densities of these sources, enabling better constraints on the contributions of the jet base, extended jet, and accretion flow, while also providing useful synergies and spectral shapes alongside existing EHT+ALMA 
observations (see Sect.~\ref{ETHER: EHT observations overview}) and other multi-wavelength (MWL) data. These GMVA observations are expected to provide better performance with the inclusion of the highly sensitive NOEMA \citep[e.g.,][]{Kim2023} and ALMA arrays \citep[e.g.,][]{Lu2023}.

\vspace{-1.0em}
\subsection{\textbf{ETHER: 230 GHz EHT+ALMA observations}}
\label{ETHER: EHT observations overview}
The results from our 43 GHz and 86 GHz VLBI imaging (see Sects.~\ref{sec:ETHER: 43 GHz VLBA Imaging} and \ref{ETHER: GMVA observations overview}), combined with data from the literature, our 230 GHz ACA and SMA surveys (Sect.~\ref{sec:230 GHz survey with ALMA Compact Array Submillimeter Array}), and our jet+ADAF modeling (Sect.\ref{sec:The sub-mm `bump' and ADAF+Jet models to predict EHT fluxes}), enabled us to select an optimal sub-sample of 27 targets for 230 GHz EHT+ALMA observations in April 2022, 2023, and 2024 with an angular resolution of $\sim$ 20 \microarc.
The 15-minute EHT+ALMA fringe tests in 2022 on five LLAGNs (\href{https://almascience.eso.org/observing/highest-priority-projects}{PID:2021.1.01156.V}, 6 hours) -- Sombrero, M84, NGC\,4261, NGC\,4278, and NGC\,5232 achieved clear detections across all baselines, and imaging of these sources was possible even with the `snapshot' nature of the observations (Ramakrishnan et al., in prep.). `Passive phasing' at ALMA was crucial for obtaining these detections.

In April 2023, EHT$+$ALMA performed fringe tests on an additional seven ETHER LLAGNs 
(NGC\,4552, NGC\,315, IERS\,0923$+$392, 2FGL\,J1635.2$+$3810,  2FGL\,J1550.7$+$0526,
BWE\,1354$+$195, TXS\,0607$-$157; \href{https://almascience.eso.org/observing/highest-priority-projects}{PID:2022.1.01055.V}, 7.2 hours). Deep imaging six-hour observations of NGC\,4261 were also conducted (\href{https://almascience.eso.org/observing/highest-priority-projects}{PID:2022.1.00520.V}).

In April 2024, EHT$+$ALMA carried out fringe tests on twelve more ETHER LLAGNs 
(NGC\,6500, 4C01.24B (J0909$+$0121), 2FGL\,J0956.9$+$2516, 2FGL\,J1159.5$+$2914, OQ208, IERS\,1741$-$038, PKS\,1936$-$15, 2FGL\,J2025$-$0736, 
OX$-$146 (J2129$-$1538), IERS\,2128$-$123, PKS\,2344$+$092)
and repeated observations on three LLAGNs (NGC\,315, IERS0738$+$313, and 2FGLJ1504.3$+$1029), which had also been observed in 2023 (\href{https://almascience.eso.org/observing/highest-priority-projects}{PID:2023.1.01139.V}, 9 hours). Deep imaging 12-hour observations of the Sombrero and M84 galaxies were also conducted (6 hours each; \href{https://almascience.eso.org/observing/highest-priority-projects}{PID:2023.1.01175.V}).

Additionally, eight more hours of EHT+ALMA imaging have been approved for M84, scheduled for April 2025 (\href{https://almascience.eso.org/observing/highest-priority-projects}{PID:2024.1.00802.V}).

The completed (2023--2024) and approved (2025) deep imaging observations of NGC\,4261, Sombrero, and M84, each for 6--8 hours, are expected to uniquely reveal the inner jet bases and potentially resolve the black hole shadows within them. The expectations based on the data sets of Sombrero and M84 are shown in detail in Figs.\ref{Sombrero-grmhd}, \ref{Sombrero-synthetic}, and \ref{M84-synthetic}, which were generated using the \href{https://github.com/achael/eht-imaging}{\texttt{eht-imaging}} package \citep{Chael2018}.

Note that tests with synthetic data, particularly with a simple geometric source models fitted with the same family of models, as considered in Figs.~\ref{Sombrero-synthetic} and~\ref{M84-synthetic}, are idealized. Successful source morphology reconstruction in such cases should be considered a necessary rather than a sufficient condition to determine the positive outcome of the actual observations. Even in the case of a GRMHD model of the source, as in Fig.~\ref{Sombrero-grmhd}, some simplifications were taken, e.g., 'passive' phasing efficiency at ALMA are not taken into account, and good weather conditions were assumed for all the sites. Narrow fit uncertainties are a consequence of underfitting a simple geometric model to complicated synthetic data. Our ongoing observational projects will give us a better understanding of the model fitting process for low signal-to-noise sources. In any case, with the inclusion of more baselines and the upgrades of the EHT sites, the EHT's imaging capabilities are expected to vastly improve in the future \citep[e.g.,][]{Johnson2023, Doeleman2023}.

\vspace{-1.0em}       
\section{Conclusions \& prospects}

In this article, we present the capabilities of the ETHER database and the identified `ETHER GS', along with SED modeling and VLBI programs, to accurately measure VLBI flux densities from SMBHs with large apparent ring sizes at high angular resolutions using VLBA, GMVA, SMA, ALMA, and EHT across frequencies from 43 to 230 GHz. The conclusions are as follows:

\begin{itemize}

\item The ETHER database provides candidate SMBHs that can potentially be `extreme' laboratories for a wide range of science goals, allowing unique tests of physics, including GR, accretion processes, jet launching, and the properties of plasma and particles.

\item VLBA imaging of 32 of the largest SMBH ring ETHER targets helped constrain the 43 GHz nuclear flux densities and revealed morphologies ranging from core-twin-jet to core-jet and compact core. This insight enabled deep imaging of this optimal sub-sample with GMVA$+$ALMA and EHT$+$ALMA to investigate their jet launching and accretion inflows at $\leq$ 100 \rg.

\item In our EHT+ALMA observations in 2022, five of the top ETHER GS galaxies - Sombrero, NGC\,4261, M84, NGC\,4278, and NGC\,5232 are clearly detected, indicating that the 230 GHz flux density from the accretion flows is significantly high.

\item Integrating 43 GHz VLBA radio flux densities with 5-arcsec 230 GHz flux densities from ACA and SMA campaigns is essential for ADAF+jet SED fitting of ETHER targets. 
By incorporating X-ray flux densities from \textit{Chandra}, \textit{eROSITA}, and our \textit{NuSTAR} and \textit{NICER} observations, along with the ETHER database, we can predict EHT flux densities through SED modeling with ADAF and ADAF+jet models, helping in EHT and ngEHT target selection.

\item  Our initial ACA and SMA survey results of 1,025 AGNs and EHT-predicted 230 GHz flux densities suggest that SMBH demographic studies with the ngEHT would benefit from the capability to image $\sim$ 10 mJy SMBHs and detect $\sim$ 1 mJy SMBHs in $\sim$ 1 hour of integration.

\item  Our completed and scheduled deep EHT+ALMA imaging of ETHER GS galaxies, combined with MWL observations, offers the potential to study SMBH masses, spins, and potentially resolved rings, thereby constraining both GR and the physics of accretion flows across a broad parameter space (SMBH mass, accretion rate, spin, jet power/type, and orientation). This will allow us to resolve or constrain the accretion flow extent (Fig. \ref{figadaf}) and search for orbiting hot spots at $\gtrsim$ 30--100 \rg\ \citep{Doeleman2009,Fish2013,Emami2023}. Monitoring these features will provide strong constraints on the GR metric and accretion physics.

\item With 11 telescopes, including passive-phased ALMA and NOEMA, 
we expect to detect sources as faint as $\sim$ 30--50 mJy.
This capability may enable us to leverage the transformational results in M87 and Sgr A* to a few other nearby AGNs with the current EHT, and 
even larger samples of AGNs with the ngEHT.

\end{itemize}

\begin{acknowledgements}
\scriptsize
The Submillimeter Array is a joint project between the Smithsonian Astrophysical Observatory and the Academia Sinica Institute of Astronomy and Astrophysics and is funded by the Smithsonian Institution and the Academia Sinica. We recognize that Maunakea is a culturally important site for the indigenous Hawaiian people; we are privileged to study the cosmos from its summit. 
This paper makes use of following ALMA data: ADS/JAO.ALMA\#2021.1.00966.S, ADS/JAO.ALMA\#2022.1.00049.S, ADS/JAO.ALMA\#2023.1.01099.S. ALMA is a partnership of ESO (representing its member states), NSF (USA), and NINS (Japan), together with NRC (Canada), MOST and ASIAA (Taiwan), and KASI (Republic of Korea), in cooperation with the Republic of Chile. The Joint ALMA Observatory is operated by ESO, AUI/NRAO, and NAOJ.
This research has made use of data obtained with the Global Millimeter VLBI Array (GMVA), which consists of telescopes operated by the MPIfR, IRAM, Onsala, Metsahovi, Yebes, the Korean VLBI Network, the Greenland Telescope, the Green Bank Observatory and the Very Long Baseline Array (VLBA). 
The VLBA and the GBT are facilities of the National Science Foundation operated under cooperative agreement by Associated Universities, Inc. The data were correlated at the correlator of the MPIfR in Bonn, Germany. We thank the EHT Collaboration for their work on enabling the extreme resolution VLBI observations. The VLBA is a facility of the National Science Foundation operated by the National Radio Astronomy Observatory under cooperative agreement with Associated Universities, Inc.
DGN acknowledges funding from ANID through Fondecyt Postdoctorado (project code 3220195) and Nucleo Milenio TITANs (project NCN2022\_002). 
SS acknowledges financial support from Millenium Nucleus NCN23\_002 (TITANs) and Comite Mixto ESO-Chile. MM acknowledges support from the Spanish Ministry of Science and Innovation through the project PID2021-124243NB-C22. This work was partially supported by the program Unidad de Excelencia Maria de Maeztu CEX2020-001058-M. 
\end{acknowledgements}


\small{
\textbf{Affiliations}
\vspace{-0.2em}
\begin{itemize}
    \item[1] Astronomy Department, Universidad de Concepci\'on, Casilla 160-C, Concepci\'on, Chile
    \item[2] Max-Planck-Institut f\"ur Radioastronomie, Auf dem H\"ugel 69, D-53121 Bonn, Germany
    \item[3] Finnish Centre for Astronomy with ESO, FI-20014 University of Turku, Finland
    \item[4] Instituto de Astrofisica de Andalucia-CSIC, Glorieta de la Astronomia s/n, E-18008 Granada, Spain
    \item[5] Center for Astrophysics \textbar\ Harvard \& Smithsonian, 60 Garden Street, Cambridge, MA 02138, USA
    \item[6] Department of Physics, McGill University, 3600 rue University, Montr\'eal, QC H3A 2T8, Canada
    \item[7] Massachusetts Institute of Technology Haystack Observatory, 99 Millstone Road, Westford, MA 01886, USA
    \item[8] Black Hole Initiative at Harvard University, 20 Garden Street, Cambridge, MA 02138, USA
    \item[9] National Astronomical Observatory of Japan, 2-21-1 Osawa, Mitaka, Tokyo 181-8588, Japan
    \item[10] Department of Space, Earth and Environment, Chalmers University of Technology, Onsala Space Observatory, SE-43992 Onsala, Sweden      
    \item[11] Institute of Astronomy and Astrophysics, Academia Sinica, 645 N. A'ohoku Place, Hilo, HI 96720, USA
    \item[12] Department of Physics and Astronomy, University of Hawaii at Manoa, 2505 Correa Road, Honolulu, HI 96822, USA   
    \item[13] Perimeter Institute for Theoretical Physics, 31 Caroline Street North, Waterloo, ON N2L 2Y5, Canada
    \item[14] Department of Physics and Astronomy, University of Waterloo, 200 University Avenue West, Waterloo, ON N2L 3G1, Canada
    \item[15] Waterloo Centre for Astrophysics, University of Waterloo, Waterloo, Canada
    \item[16] Steward Observatory and Department of Astronomy, University of Arizona, 933 N. Cherry Ave., Tucson, AZ 85721, USA
    \item[17] Data Science Institute, University of Arizona, 1230 N. Cherry Ave., Tucson, AZ 85721, USA
    \item[18] Program in Applied Mathematics, University of Arizona, 617 N. Santa Rita, Tucson, AZ 85721, USA
    \item[19] Department of Astrophysics, Institute for Mathematics, Astrophysics and Particle Physics (IMAPP), Radboud University, P.O. Box 9010, 6500 GL Nijmegen, The Netherlands
    \item[20] Institut f\"ur Theoretische Physik, Goethe-Universit\"at Frankfurt, Max-von-Laue-Stra\ss e 1, D-60438 Frankfurt am Main, Germany
    \item[21] Institut f\"ur Theoretische Physik und Astrophysik, Universit\"at W\"urzburg, Emil-Fischer-Str.\ 31, Germany
    \item[22] Trottier Space Institute at McGill, 3550 rue University, Montr\'eal, QC H3A 2A7, Canada 
    \item[23] Mizusawa VLBI Observatory, National Astronomical Observatory of Japan, 2-12 Hoshigaoka, Mizusawa, Oshu, Iwate 023-0861, Japan
    \item[24] Department of Astronomical Science, The Graduate University for Advanced Studies (SOKENDAI), 2-21-1 Osawa, Mitaka, Tokyo 181-8588, Japan
    \item[25] Institute for Astrophysical Research, Boston University, 725 Commonwealth Ave., Boston, MA 02215, USA
    \item[26] National Centre for Radio Astrophysics (NCRA) - Tata Institute of Fundamental Research (TIFR), S. P. Pune University Campus, Ganeshkhind, 411007 Pune, India 
    \item[27] Department of Astronomy, Kyungpook National University, 80 Daehak-ro, Buk-gu, Daegu, 41566, Republic of Korea 
    \item[28] Institute of Astronomy and Astrophysics, Academia Sinica, 11F of Astronomy-Mathematics Building, AS/NTU No. 1, Sec. 4, Roosevelt Rd., Taipei 106216, Taiwan, R.O.C. 
    \item[29] Graduate School of Science and Technology, Niigata University, 8050 Ikarashi 2-no-cho, Nishi-ku, Niigata 950-2181, Japan 
    \item[31] Shanghai Astronomical Observatory, Chinese Academy of Sciences, 80 Nandan Road, Shanghai 200030, People's Republic of China
    \item[32] Key Laboratory of Radio Astronomy, Chinese Academy of Sciences, Nanjing 210008, People's Republic of China
    \item[33] Anton Pannekoek Institute for Astronomy, University of Amsterdam, Science Park 904, 1098 XH, Amsterdam, The Netherlands
    \item[34] Gravitation and Astroparticle Physics Amsterdam (GRAPPA) Institute, University of Amsterdam, Science Park 904, 1098 XH Amsterdam, The Netherlands
    \item[35] Department of Astronomy, University of Massachusetts, Amherst, MA 01003, USA
    \item[36] Physics Department, Washington University, CB 1105, St. Louis, MO 63130, USA
    \item[37] School of Space Research, Kyung Hee University, 1732, Deogyeong-daero, Giheung-gu, Yongin-si, Gyeonggi-do 17104, Republic of Korea
    \item[38] School of Physics, Georgia Institute of Technology, 837 State St NW, Atlanta, GA 30332, USA
    \item[39] Aalto University Department of Electronics and Nanoengineering, PL 15500, FI-00076 Aalto, Finland
    \item[40] Aalto University Mets\"ahovi Radio Observatory, Mets\"ahovintie 114, FI-02540 Kylm\"al\"a, Finland
    \item[41] Leiden Observatory, Leiden University, 9513 RA Leiden, The Netherlands
    \item[42] National Radio Astronomy Observatory, 520 Edgemont Road, Charlottesville
    \item[43] Millennium Nucleus on Transversal Research and Technology to Explore Supermassive Black Holes (TITANS)
    \item[44] Instituto de F\'isica y Astronom\'ia, Universidad de Valpara\'iso, Gran Breta\~na 1111, Valpara\'iso, Chile
    \item[45] D\'epartement de Physique, Universit\'e de Montr\'eal, Succ. Centre-Ville, Montr\'eal, QC, H3C 3J7, Canada
    \item[46] Institute of Space Sciences (ICE, CSIC), Campus UAB, Carrer de Magrans, 08193 Barcelona, Spain
    \item[47] Institut d\textquotesingle{}Estudis Espacials de Catalunya (IEEC), Edifici RDIT, Campus UPC, 08860 Castelldefels (Barcelona), Spain
    \item[48] Department of Astronomy, Yale University, 219 Prospect Street, New Haven, CT 06511, USA
    \item[49] Department of Physics, Yale University, 217 Prospect Street, New Haven, CT 06511, USA
    \item[50] Instituto de Astrof\'isica and Centro de Astroingenier\'ia, Facultad de F\'isica, Pontificia Universidad Cat\'olica de Chile, Casilla 306, Santiago 22, Chile
    \item[51] National Institute of Technology, Hachinohe College, 16-1 Uwanotai, Tamonoki, Hachinohe City, Aomori 039-1192, Japan
    \item[52] Institute for Cosmic Ray Research, University of Tokyo, Chiba 277-8582, Japan
    \item[53] Physics Department, National Sun Yat-Sen University, No. 70, Lien-Hai Road, Kaosiung City 80424, Taiwan, R.O.C.
    \item[54] Tuorla Observatory, Department of Physics and Astronomy, University of Turku, Finland
    \item[55] Departamento de Matem\'atica da Universidade de Aveiro and Centre for Research and Development in Mathematics and Applications (CIDMA), Campus de Santiago, 3810-193 Aveiro, Portugal
    \item[56] Department of Physics, National Taiwan University, No. 1, Sec. 4, Roosevelt Rd., Taipei 106216, Taiwan, R.O.C.
    \item[57] Instituto Nacional de Astrof\'isica, \'Optica y Electr\'onica. Apartado Postal 51 y 216, 72000. Puebla, Pue., M\'exico
    \item[58] Department of Physics, National Taiwan Normal University, No. 88, Sec. 4, Tingzhou Rd., Taipei 116, Taiwan, R.O.C.
    \item[59] Center of Astronomy and Gravitation, National Taiwan Normal University, No. 88, Sec. 4, Tingzhou Road, Taipei 116, Taiwan, R.O.C.
    \item[60] Institute of Space and Astronautical Science, Japan Aerospace Exploration Agency, Japan
    \item[61] CSIRO, Space and Astronomy, PO Box 1130, Bentley, WA 6151, Australia
    \item[62] Manly Astrophysics, 15/41-42 East Esplanade, Manly NSW 2095, Australia
    \item[63] Department of Physics, Ulsan National Institute of Science and Technology (UNIST), Ulsan 44919, Republic of Korea
    \item[64] Department of Physics and Astronomy, University of Mississippi, 108 Lewis Hall, University, Mississippi 38677-1848, USA.
\end{itemize}
}

\end{document}